\documentclass{llncs}
\usepackage[linesnumbered,ruled]{algorithm2e}
\usepackage{graphicx}
\usepackage{amsmath}
\usepackage{multirow}
\usepackage{scrextend}
\usepackage{color}
\usepackage{tabularx}
\usepackage{graphicx,wrapfig,lipsum}

\newcommand{\name}{$\mathsf{DyPerm}$}

\begin{document}

\title{DyPerm: Maximizing Permanence for Dynamic Community Detection\thanks{{\color{blue}The work is accepted in 22nd Pacific-Asia Conference on Knowledge Discovery and Data Mining (PAKDD), 2018}}}
%
%
\author{Prerna Agarwal$^1$, Richa Verma$^2$, Ayush Agarwal$^3$, Tanmoy Chakraborty$^4$}
\institute{$^1$IBM Research, India, $^2$TCS Innovation Lab, India, $^{3,4}$IIIT Delhi, India\\
$^1$prernaagarwal@in.ibm.com;\{$^2$richa15054,$^3$ayush14029,$^4$tanmoy\}@iiitd.ac.in
}
\authorrunning{Prerna Agarwal et al.} 
%
%

\maketitle              

\begin{abstract}
In this paper, we propose \name, the first dynamic community detection method which optimizes  a novel community scoring metric, called {\em permanence}.
\name~incrementally modifies the community structure by updating those communities where the editing of nodes and edges has been performed, keeping the rest of the network unchanged. We present strong theoretical guarantees to show how/why mere updates on the existing community structure leads to permanence maximization in dynamic networks, which in turn decreases the computational complexity drastically. Experiments on both synthetic and six real-world networks with given ground-truth community structure show that \name~achieves (on average) 35\% gain in accuracy (based on NMI) compared to the best method among four baseline methods. \name~also turns out to be $15$ times faster than its static counterpart.\footnote{\label{footnote}For anonymized code and datasets, please visit \url{https://tinyurl.com/dyperm-code}.}
\end{abstract}
\vspace{-10mm}
\section{Introduction}
\vspace{-3mm}
Last one decade has witnessed tremendous advancement in the detection and analysis of community structure (densely connected groups containing homogeneous nodes) in different types of networks \cite{Chakraborty:2017}. So far, major research has concentrated on detecting communities from static networks \cite{Fortunato201075}. However, today's     
real-world networks, especially most of the social networks, are not always static -- networks such as Facebook, Twitter are evolving heavily and expanding rapidly in terms of both
size and complexity over time. This has recently led to turn the research focus from static networks to dynamic networks (where nodes and edges are added/deleted continuously) \cite{Cazabet2014}.  The evolving nature of network structure raises several new challenges to traditional community detection methods -- on one hand, the new community structure obtained due to certain changes in the network structure should not be drastically different from that in the previous time-stamp; on the other hand, the algorithm needs to guarantee that the communities has a dynamic adaptability to deal with the dynamic events.  

Existing research on dynamic community detection either run static community detection method on different snapshot of the networks \cite{MITRA20121041} and then correlate the community structures in two consecutive time-stamps, or adopt standard community goodness metrics such as modularity \cite{Chakraborty:2017} and optimize them to obtain final communities \cite{7395763,abs-1305-2006}. In this paper, we propose \name, {\em the first dynamic community detection method that adopts an effective community goodness metric, called} ``permanence'' \cite{Chakraborty:2014,Chakraborty:2016,0002KGMB16} and  {\em optimizes it to incrementally detect the community structure}. The benefits of adopting permanence as an optimization function are two-fold: (i) Permanence, being a local vertex-centric metric (as opposed to the global network-centric metrics such as modularity, conductance), allows us to reassign communities to only those nodes whose associated topological structure has changed, and guarantees that the remaining nodes do not affect the optimization. This leads to very low computing complexity in updating the community structure when the network changes dynamically. (ii) Incremental changes in the local portion of the community structure guarantee that the resultant communities are highly correlated with that in the previous time-stamp. We present theoretical justifications why/how mere changes in the community structure lead to maximize permanence.

We experiment with both synthetic and six real-world dynamic networks with known ground-truth community structure. A thorough comparative evaluation with four state-of-the-art baseline methods shows that \name~significantly outperforms all the baselines across different networks -- \name~achieves up to 35\% improvement in terms of Normalized Mutual Information  (NMI) w.r.t. the best baseline method. Moreover, \name~tunrs out to be extremely fast, achieving up to 15 times speedup w.r.t. its static counterpart. In short, \name~is a fast and accurate dynamic community detection method.

\vspace{-4mm}
\section{Related work}
\vspace{-4mm}
Community detection has been extensively studied  in last one decade mostly for the static networks (see \cite{Chakraborty:2017,Fortunato201075,Xie:2013} for comprehensive reviews). However, due to the enormous growth of the network size and the evolving nature of the network structure, people turned their focus from static network to dynamic network. Major research on dynamic community detection can be divided into three categories \cite{Cazabet2014}: (i) {\bf traditional clustering} where a static community detection method is applied to different snapshots of the dynamic networks \cite{MITRA20121041}; (ii) {\bf evolutionary clustering} \cite{Chakrabarti:2006,LiuHSWQL15} where clustering at a particular time-stamp should be similar to the clustering of the previous time-stamp and should accurately reflect the data arriving during that time, and  (iii) {\bf incremental clustering} \cite{abs-1305-2006,7395763} where given the clustering result of the initial snapshot, it incrementally modifies clusters based on every occurrence of an event in the network. Modularity \cite{Chakraborty:2017}, a well-studied goodness metric for static communities, has recently been adopted for dynamic community detection \cite{Aktunc:2015,DBLP:journals,journal.pone.0091431,Bansal2011}.  QCA \cite{5935045,journal.pone.0091431} is another such method which adopts modularity to identify and trace dynamic communities.  Shang et al. \cite{ShangLXCMFW14} proposed GreMod which  first uses Louvain algorithm \cite{louvain} to detect the initial community structure, and then applies incremental updating strategies to track the dynamic communities. They further proposed LBTR \cite{shang2016targeted} which uses machine learning classifiers to predict the vertices that need to be inspected for community assignment revision.

Our proposed method \name~is an incremental clustering method. It differs from the earlier methods in terms of at least two aspects: (i) \name~is the first dynamic community detection method which optimizes ``permanence'' \cite{Chakraborty:2014}, a novel metric that has been shown to be very effective compared to other state-of-the-art optimization functions (such as modularity, conductance etc.) for static communities \cite{Chakraborty:2014,Chakraborty:2017}; (ii) permanence, being a local vertex-centric metric (as opposed to the global metrics such as modularity), enables \name~to inspect only a small portion of the network (which has recently been changed), thus leading to obtain accurate communities with minimum changes \cite{Chakraborty17}.   

\if{0}
\begin{figure}[!t]
\includegraphics[width=\columnwidth]{toyexample.png}
\vspace{-5mm}
\caption{Toy examples demonstrating (a) the calculation of permanence, (b) intra-community edge (dotted red), and (c) inter-community edge (dotted red).}\label{example}
\vspace{-5mm}
\end{figure}

\fi

\vspace{-4mm}
\section{Methodology}\label{sec:method}
\vspace{-4mm}
\name~({\bf Dy}namic community Detection by maximizing {\bf Perm}anence) is an incremental method which maximizes a vertex-centric community scoring metric, called {\em permanence} \cite{Chakraborty:2014}. In this section, we start by providing a brief idea of permanence, followed by a detailed description of our proposed method.

\noindent\textbf{\underline{Permanence}:} It measures the extent to which a vertex $v$ remains consistent inside a community $c$ based three factors \cite{Chakraborty:2014}: (i) {\em $v$'s internal connectedness}, measured by the ratio of its internal neighbors inside $c$, $I(v)$ to its degree $d(v)$; (ii) {\em $v$'s cohesiveness}, indicating how connected its internal neighbors are and measured by  $C_{in}(v)=\frac{E_{neig}(v)}{\binom {I(v)} {2}}$, the ratio of actual number of connections among its internal neighbors $E_{neig}(v)$ to the total number of possible connections among them; and (iii) {\em $v$'s external pull}, measured by the maximum number of external connections of $v$ to any of the external communities $E_{max}(v)$. These three factors are suitably combined to obtain permanence of $v$ as follows:
\begin{equation}\label{eq:permanence}
Perm(v) = [\frac{I(v)}{E_{max}(v)} \times \frac{1}{d(v)}]-[1-C_{in}(v)]
\end{equation}

\begin{wrapfigure}{r}{4cm}
\vspace{-10mm}
\includegraphics[width=4cm]{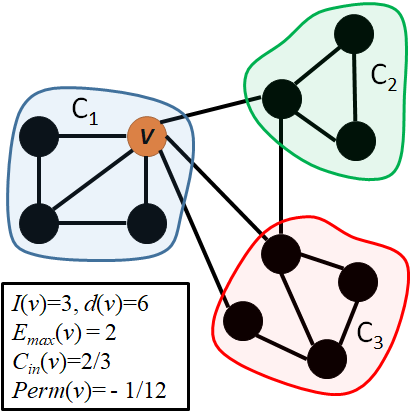}
\caption{A toy example showing permanence of vertex $v$. $C_1$, $C_2$ and $C_3$ are three communities.}\label{toy}
\vspace{-7mm}
\end{wrapfigure}

Figure \ref{toy} shows an illustrative example to calculate permanence of a vertex. If $E_{max}=0$, then  $Perm(v)=\frac{I(v)}{d(v)}$.
Given a network $G(V,E)$ and its community structure $C$, permanence of the graph is obtained by $Perm(G)=\frac{1}{|V|} \sum_{v \in V} Perm(v)$. $Perm(G)$ always ranges between $-1$ (indicating weak community structure) to $1$ (indicating strong community structure). We chose permanence as our objective function for two reasons: 
(i) it is a local vertex-centric metric, which enables us to inspect only the changes happened in a local portion of the network, instead of looking at all the changes as a whole, and (ii)  it was already shown to outperform many well-studied local and global metrics such as modularity, conductance, SPart, significance etc. on different static networks (see a detailed survey in \cite{Chakraborty:2017}).      

\noindent\textbf{\underline{Dynamic network}:}
A dynamic network $\mathcal{G}(E,V)$ can be conceptualized by a time evolving process where the underlying network is continuously updated over time by either inserting or removing nodes/edges. 
Therefore, the atomic events can be of  following types:  

\begin{itemize}
\item $newNode(V\cup u)$: A node $u$ is added to the network. It may or may not have one or more associated edges. 
\item $removeNode(V\setminus u)$: A node $u$ is removed from the network along with its associated edges. 
\item $newEdge(E\cup e)$: A new edge $e$ is added between two existing nodes in the network. 
\item $removeEdge(E \setminus e)$: An existing edge $e$ is removed from the network. 
\end{itemize}

Therefore, the dynamic network $\mathcal{G}$ can be expressed as a collection of $t$ static snapshots $\mathcal{G}=\{ G_0, G_1, G_2, \cdots, G_t\}$, where $G_{i+1}=G_i \cup \Delta G_{i}$ indicates the static snapshot of $\mathcal{G}$ at $(i+1)^{th}$ time-stamp. $G_{i+1}$ has evolved from $G_i$ due to $\Delta G_{i}$ change in $G_i$, where $\Delta G_{i}$ is one of the four atomic events mentioned above.
\vspace{-5mm}

\if{0}
The below table describes the notation used.
\begin{table}[]
\centering
\begin{tabular}{|l|l|}
\hline
Notation & Description                  \\ \hline
$Perm^{C}(u)$  & Permanence of a node $u$ in community C   \\ \hline
$Perm^{'C}(u)$  & New permanence of a node $u$ in community C\\ \hline
$Comm(u)$       & Community of node $u$        \\ \hline
$C_t$       & Current community structure  \\ \hline
$C_{t+1}$     & Updated community structure  \\ \hline
$e_{u,v}$     & Edge between nodes $u$ and $v$   \\ \hline
$I(u)$ & Number of internal neighbors of u \\   \hline
$I^{'}(u)$ & Number of updated internal neighbors of u \\   \hline
$E_{max}(u)$ & Maximum number of neighbors of $u$ contributed by its  neighboring communities\\ \hline
$d(u)$ & Degree of node $u$\\ \hline
$C_{in}(u)$ & Internal clustering coefficient\\ \hline
\end{tabular}
\caption{Notations used}
\label{my-label}
\end{table}
\fi

\subsection{The \name~algorithm}
\vspace{-3mm}
\name~requires the community structure $C_0$ (referred as {\em base community structure})  of the initial snapshot $G_0$, which can be obtained by running a static community detection method on $G_0$ or from an oracle who knows the ground-truth community structure of $G_0$. Depending upon the atomic event which causes the change in the network structure, \name~executes one of the following routines in order to maximize permanence:

\noindent\textbf{(A) \underline{Addition of a new node:}} When a new node $u$ is added into the network (i.e., case: $newNode(V\cup u))$, two scenarios may arise (see Algorithm 1):\\
$\bullet$ \noindent \textbf{Case A.1:} $u$ does not have any associated edges. It then forms a new singleton community containing only itself.\\
$\bullet$ \noindent \textbf{Case A.2:} $u$ has  more than one associated edges. Adding $u$ can be approached as inserting edge(s) associated with $u$, one by one, if the order of edge addition does not affect the final community structure (shown in Proposition \ref{prop1}).  

\begin{proposition}\label{prop1}
The order in which the edges (both intra- and inter-community) associated with a node are inserted, is immaterial for permanence maximization.
\end{proposition}
See supplementary \cite{si} for the proof.

\if{0}
\begin{proof}
There are two types of edges associated with $u$: (i) intra-community edges whose both end nodes are present inside $C_u$, and (ii)  inter-community edges whose one end node is inside $C_u$, other is outside $C_u$.   
In the case of adding an intra-community edge, $I(u)$ and $d(u)$ in Eq. \ref{eq:permanence}  increase by 1. Since, the increase is the same, we say that the order in which intra-community edges of a node are added does not matter.
In the case of adding an inter-community edge, the value of permanence decreases or remains same depending upon the change in $E_{max}(u)$. But, as shown in Algorithm 5, $Perm(u)$ is checked after moving it to $v$'s community and after moving its neighbors, one by one, till the permanence value keeps increasing. The same steps are followed for $v$, after moving it to $u$'s community. The order in which the edges are inserted does not matter because in any case, the algorithm will displace the nodes and their neighbors so as to maximize the permanence value of each community, separately.
\end{proof}

\fi

\begin{figure}[!t]
\centering
\includegraphics[width=\columnwidth]{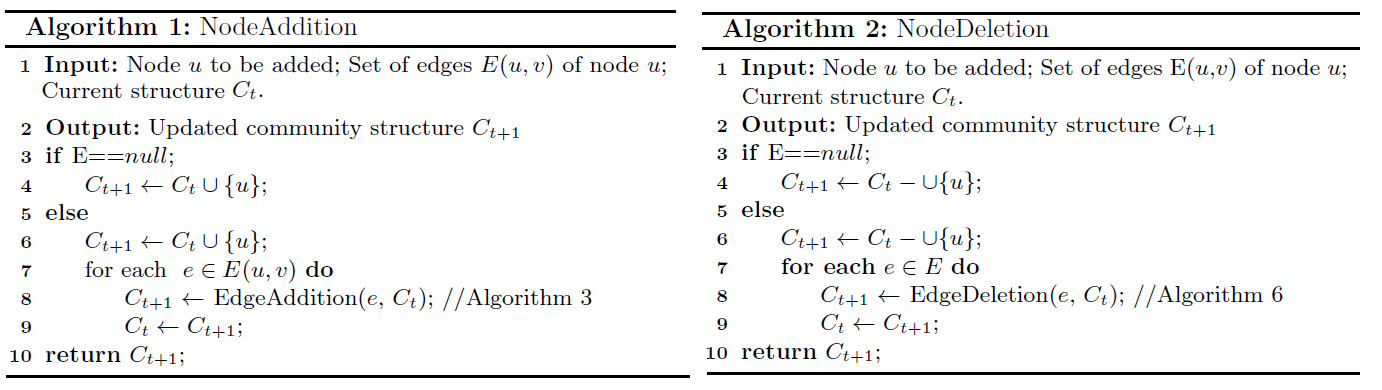}
\vspace{-10mm}
\end{figure}

\if{0}


\setlength{\textfloatsep}{0pt}
\begin{algorithm}\label{algo:node_add}\small
  \textbf{Input:} Node $u$ to be added; Set of edges $E(u,v)$ of node $u$; Current structure $C_t$.\\
   \textbf{Output:} Updated community structure $C_{t+1}$ \\
  \SetAlgoLined
  \textbf{if} E==$null$;\\
\Indp $C_{t+1} \leftarrow C_{t} \cup \{u\}$;\\

\Indm \textbf{else}\\
\Indp $C_{t+1} \leftarrow C_{t} \cup \{u\}$;\\
{for each } $e\in E(u,v)$ \textbf{do}\\
\Indp $C_{t+1}$ $\leftarrow$ EdgeAddition($e$, $C_t$);    //Algorithm 3 \\
$C_t \leftarrow C_{t+1}$;\\ 

   \Indm \Indm \textbf{return} $C_{t+1}$; \\
  \caption{NodeAddition}
\end{algorithm}
%

\fi

\noindent\textbf{(B) \underline{Removal of an existing node:}} When an existing node $u$ present in community $C_u$ is removed (i.e., case: $removeNode(V \setminus u)$), its associated edges are also detected (see Algorithm 2). Therefore, node removal can be handled by deleting the associated edges, one by one, if the order of edge deletion does not affect the final community structure (see Proposition \ref{prop2}).

\begin{proposition}\label{prop2}
The order in which the edges (both intra- and inter-community) associated with a node are deleted, is immaterial for permanence maximization.
\end{proposition}
See Supplementary \cite{si} for the proof. 
\if{0}
\begin{proof}
In the case of deleting intra-community edges, the order is immaterial because with the deletion of every internal edge connected with $u$, both $I(u)$ and $d(u)$ decrease by one. 
In the case of deleting inter-community edges, the order of deletion is immaterial. For every edge to an external neighbor of a node $u$ that is deleted, $d(u)$ will decrease and $E_{max}(u)$ may decrease or remains same.   $C_{in}(u)$ remains unchanged. Therefore, $Perm(u)$ increases. So each deletion is favorable for $u$. Even if the order of deletion is changed, the resultant permanence value will increase. 
\end{proof}
\fi

\if{0}
\vspace{-5mm}
\begin{algorithm}\small
  \textbf{Input:} Node $u$ to be added; Set of edges E($u$,$v$) of node $u$; Current structure $C_t$.\\
   \textbf{Output:} Updated community structure $C_{t+1}$ \\
  \SetAlgoLined
  \textbf{if} E==$null$;\\
\Indp $C_{t+1} \leftarrow C_{t} - \cup \{u\}$;\\
\Indm \textbf{else}\\
\Indp $C_{t+1} \leftarrow C_{t} - \cup \{u\}$;\\
{\bf for each} $e \in E$ \textbf{do}\\
\Indp $C_{t+1}$ $\leftarrow$ EdgeDeletion($e$, $C_t$);    //Algorithm 6\\
$C_t \leftarrow C_{t+1}$;\\

   \Indm \Indm \textbf{return} $C_{t+1}$; \\
  \caption{NodeDeletion}
\end{algorithm}
\vspace{-5mm}
\fi

\noindent\textbf{(C) \underline{Addition of a new edge:}}
Let us consider adding an edge $e_{u,v}$ between two existing nodes $u$ and $v$. There are two possible cases (See Algorithm 3):\\
$\bullet$ \noindent\textbf{Case C.1: Addition of an intra-community edge:} Both $u$ and $v$ belong to the same community $C$. Accordingly to Proposition \ref{prop3}, addition of $e_{u,v}$ will increase the permanence value of the entire network, and the community will not split into smaller communities (See Algorithm 4). 
\begin{proposition}\label{prop3}
If $C$ is a community in the current snapshot of $\mathcal{G}$, then adding any intra-community edge to $C$ does not split it into smaller communities.
\end{proposition}
See Supplementary \cite{si} for the proof. 

\if{0}
\begin{proof}
We prove this by showing that the contribution of $u$ in $C$ is more than that when $C$ splits.
Permanence of $u$ after edge addition is:
$Perm^{C}(u) = [\frac{I(u)+1}{E_{max}(u)} \times\frac{1}{d(u)+1}] - [1- C_{in}(u)]$.
Since permanence is a vertex-centric metric, we have to look at the effect it has on neighboring communities of $u$ as well. Let both $u$ and $v$ belong to community $C$. If $C$ splits into two smaller communities, one containing $u$ and the other containing $v$, the new $E_{max}(u)$, i.e.,$E_{max}^{'}(u)$ ranges from $E_{max}(u)$ to $I(u)$ since it can be equal to, at most, the number of internal neighbors of $u$. Also, the new number of internal neighbors of $u$, i.e., $I^{'}(u)$ ranges from $0$ (where only $u$ gets disconnected and forms a new community) to $I(u)-1$ (when just one neighbor of $u$ goes away). Therefore, it further generates 4 sub-cases:

\textbf{Case C.1.1: $E_{max}^{'}(u) = E_{max}(u)$ and  $I^{'}(u) = 0 $}. In this case, the modified value of $Perm^{C}(u)$, i.e., $Perm^{'C}(u) = 0 - [1- C_{in}(u)] \leq 0 $.  As per the definition of Permanence of a vertex, it cannot be negative. Therefore, this case cannot occur.

\textbf{Case C.1.2: $E_{max}^{'}(u) = E_{max}(u)$ and  $I^{'}(u) = I(u) -1$}. In this case, $Perm^{'C}(u) =  [\frac{I(u)-1}{E_{max}(u)}\times \frac{1}{d(u)+ 1}]- [1- C_{in}(u)] < Perm^{C}(u)$. This case does not occur because it does not result in Permanence getting maximized. 

\textbf{Case C.1.3: $E_{max}^{'}(u) = I(u)$ and  $I^{'}(u) = 0 $}. In this case, $Perm^{'C}(u) = 0 - [1- C_{in}(u)] < 0 $. As per the definition of Permanence of a vertex, it cannot be negative. Therefore, this case is not possible.

\textbf{Case C.1.4: $E_{max}^{'}(u) = I(u)$ and  $I^{'}(u) = I(u)-1 $}.
In this case, $Perm^{'C}(u) = \frac{I(u)-1}{I(u)} \frac{1}{d(u)+ 1}- [1- C_{in}(u)] < Perm^{C}(u)$, since $I(u)> E_{max}(u)$. This case does not occur because it does not result in Permanence getting maximized.

Therefore, the current network structure remains intact in this case. 
\end{proof}
\fi

\noindent$\bullet$  \textbf{Case C.2: Addition of an inter-community edge:} Let $e_{u,v}$ be the edge connecting communities $C_u$ and $C_v$.
Its presence could make either $u$ or $v$ leave its current community and join the new community (See Algorithm 5). Also, if $u$ or $v$ decides to change its membership, it can advertise its new community to all its neighbors and some of them might eventually want to change their memberships as a consequence. We first move $u$ to its new community and consequently let its internal neighbors (both direct and indirect) determine their best modules to join in, using an algorithm similar to breadth first search. Similar steps are followed for $v$ after moving it to its new community. Overall permanence for both the communities, $C_u$ and $C_v$ are calculated, once before changing the communities of $u$ and $v$ (lines 8-9, Algorithm 5), then after moving $u$ and its neighbors, recursively to $C_v$. Finally,  permanence of the two communities is computed again after moving $v$ and its neighbors, recursively, to $C_u$ (lines 13-23, Algorithm 5). The neighbors of $u$ (and then $v$) are moved recursively to the other community till the move results in an increase in permanence of that node. The set of the moves that maximizes the overall permanence of the communities is finally accepted to determine the new community structure. 

\begin{figure}[!t]
\centering
\includegraphics[width=\columnwidth]{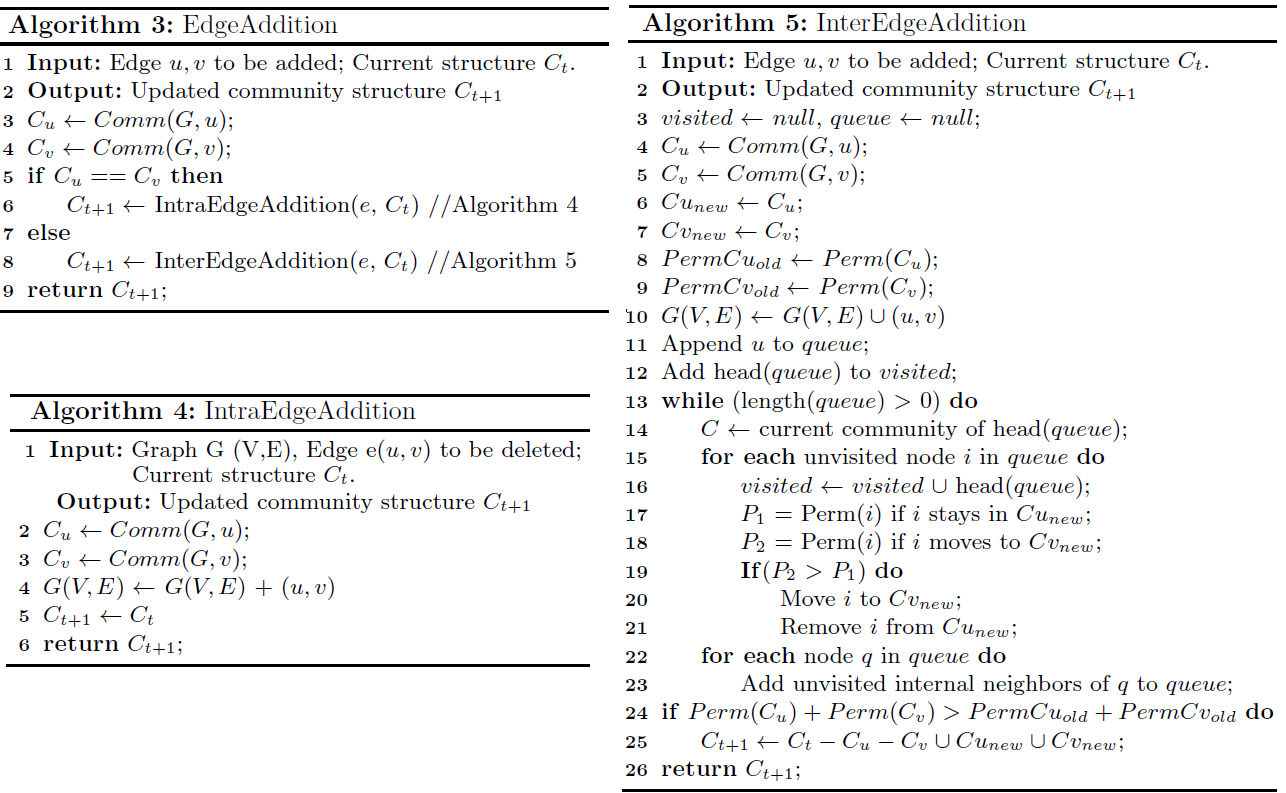}
\vspace{-10mm}
\end{figure}

\if{0}
\vspace{-8mm}
\begin{algorithm}
\textbf{Input:} Edge {$u, v$} to be added; Current structure $C_t$.\\
  \textbf{Output:} Updated community structure $C_{t+1}$ \\
  $C_u \leftarrow Comm(G,u)$;\\
  $C_v \leftarrow Comm(G,v)$;\\
  \SetAlgoLined
  \textbf{if} $C_u == C_v$ \textbf{then}\\
  \Indp $C_{t+1} \leftarrow$ IntraEdgeAddition($e$, $C_t$)  //Algorithm 4\\
  \Indm \textbf{else}\\
  \Indp $C_{t+1} \leftarrow$ InterEdgeAddition($e$, $C_t$)  //Algorithm 5\\
  \Indm \textbf{return} $C_{t+1}$; \\
  \caption{EdgeAddition}
  \end{algorithm}  
  
\begin{algorithm}

  \textbf{Input:} Graph G (V,E), Edge e($u, v$) to be deleted; Current structure $C_t$.
\textbf{Output:} Updated community structure $C_{t+1}$ \\
  $C_u \leftarrow Comm(G,u)$;\\
  $C_v \leftarrow Comm(G,v)$;\\
  \SetAlgoLined
  $G(V,E)$ $\leftarrow$ $G(V,E)$ + ($u,v$) \\
 $C_{t+1} \leftarrow C_{t}$\\
 \textbf{return} $C_{t+1}$; \\
  \caption{IntraEdgeAddition}
\end{algorithm}
\begin{algorithm}
\textbf{Input:} Edge {$u, v$} to be added; Current structure $C_t$.\\
  \textbf{Output:} Updated community structure $C_{t+1}$ \\
  \SetAlgoLined
  $visited$ $\leftarrow$ $null$, $queue$ $\leftarrow$ $null$; \\
  $C_u \leftarrow Comm(G,u)$;\\
  $C_v \leftarrow Comm(G,v)$;\\
  $Cu_{new} \leftarrow C_u$;\\
  $Cv_{new} \leftarrow C_v$;\\
  $PermCu_{old}$ $\leftarrow$ $Perm(C_u)$; \\
  $PermCv_{old} \leftarrow$ $Perm(C_v)$; \\
  $G(V,E)$ $\leftarrow$ $G(V,E) \cup  (u,v)$ \\
  Append $u$ to $queue$; \\
  Add head($queue$) to $visited$;\\
  \textbf{while} (length($queue$) $>$ 0) \textbf{do} \\
      \Indp $C$ $\leftarrow$ current community of head($queue$);  \\
            \textbf{for each} unvisited node $i$ in $queue$ \textbf{do} \\
      \Indp $visited$ $\leftarrow$ $visited$ $\cup$ head($queue$);\\
            $P_1$ = Perm($i$) if $i$ stays in $Cu_{new}$;\\
            $P_2$ = Perm($i$) if $i$ moves to $Cv_{new}$;\\
            \textbf{If}($P_2$ $>$ $P_1$) \textbf{do} \\
            \Indp Move $i$ to $Cv_{new}$; \\
            Remove $i$ from $Cu_{new}$;\\
      \Indm \Indm \textbf{for each} node $q$ in $queue$ \textbf{do}\\
      \Indp Add unvisited internal neighbors of $q$ to $queue$; \\ 
     \Indm \Indm \textbf{if} $Perm(C_u) + Perm(C_v)>PermCu_{old} + PermCv_{old}$ \textbf{do}\\
   \Indp $C_{t+1} \leftarrow C_t - C_u - C_v \cup Cu_{new} \cup Cv_{new}$;\\
   \Indm \textbf{return} $C_{t+1}$; \\
  \caption{InterEdgeAddition}
\end{algorithm}
\fi
\noindent\textbf{(D) \underline{Deletion of an existing edge:}}
Let us consider the deletion of an edge $e_{u,v}$ connecting $u$ and $v$ which are a part of existing network. There are total 3 possible cases (Algorithm 6):\\
$\bullet$ \noindent\textbf{Case D.1: Single edge connecting only $u$ and $v$:}
In this case, $u$ is only connected to $v$, and $v$ is only connected to $u$.
Let $Perm^{C}(u)$ and $Perm^{'C}(u)$ be the permanence of $u$ before and after the edge removal respectively.
If $E_{max}(u)=0$ and $d(u)< 3$, permanence of $u$ is calculated as
$Perm^{C}(u)= \frac{I(u)}{d(u)}$  (as mentioned in the beginning of Section \ref{sec:method}).
Therefore, $Perm^{C}(u) = 1$ and $Perm^{'C}(u) = 0$, i.e., the permanence value decreases. Similarly, the permanence value of $v$ will also decrease. And $u$ and $v$ form their own singleton communities. This case is handled in Algorithm 8.\\
$\bullet$ \noindent\textbf{Case D.2: Node $v$ has unit degree, i.e., $d(v)=1$}: In this case, $v$ has only one neighbor in the entire network, and $u$ can have more than one neighbors (See Algorithm 8). There are further two sub-cases:

\textbf{Case D.2.1: $u$ and $v$ belong to two different communities $C_{u}$ and $C_{v}$, respectively:} 
There are further two sub cases as permanence of $u$ is dependent upon $E_{max}(u)$, and it can be due to either $C_v$ or some other community. 

\textbf{Case D.2.1.1: $C_{v}$ is responsible for $E_{max}(u)$.}
Here, the new $E_{max}(u)$ i.e., $E_{max}^{'}(u) < E_{max}(u)$ as the edge is deleted and one neighbor goes, while everything else remains constant. The new $Perm^{C_{u}}(u)$ i.e.,  $Perm^{'C_{u}}(u) > Perm^{C_{u}}(u) $.
Permanence of $v$ remains 0 before and after edge deletion as $I(u)$ = 0. 

\textbf{Case D.2.1.2: $C_{v}$ is not responsible for $E_{max}(u)$:} In this case, the permanence values of both $u$ and $v$ will increase because the new degree of $u$ i.e., $d^{'}(u) = d(u)-1$ and the new degree of $v$ i.e., $d^{'}(v) = d(v)-1$ has decreased. Therefore, permanence increases as everything else remains constant.

\textbf{Case D.2.2: Both $u$ and $v$ belong to the same community:}
Let us assume that both $u$ and $v$ belong to community C. $Perm^{C}(v)$ = 1, and after deleting the edge, new permanence i.e., $Perm^{'C}(v) = 0$.
$Perm^{C}(u) = \frac{I(u)}{E_{max}(u)} \frac{1}{d(u)} - (1 - C_{in}(u))$.
The new degree of $u$ becomes $d^{'}(u) = d(u)-1$ and the new $I(u)$ becomes $I^{'}(u) = I(u)-1$, therefore, $Perm^{'C}(u) = \frac{I(u)-1}{E_{max}(u)} \frac{1}{d(u)-1} - (1 - C_{in}^{'}(u))$. And, $C_{in}^{'}(u) < C_{in}(u)$ because $I(u)$ has decreased; therefore $Perm^{'C}(u) < Perm^{C}(u)$.
Algorithm 7 handles this case.\\ 
$\bullet$ \noindent\textbf{Case D.3: $u$ and $v$ belong to communities $C_{u}$ and $C_{v}$, respectively and degrees of $u$ and $v$ are greater than 1:} (See Algorithm 8).
There are further four sub-cases: 

\textbf{Case D.3.1: $C_{v}$ is responsible for $E_{max}(u)$, but $C_{u}$ is not responsible for $E_{max}(v)$:}
The new $d(u)$ i.e., $d^{'}(u) < d(u)$ and the new $E_{max}(u)$ i.e., $E_{max}^{'}(u) < E_{max}(u)$ because one edge goes away. $I(u)$ remains the same. Therefore, the new permanence of $u$ i.e., $Perm^{'C_{u}}(u) > Perm^{C_{u}}(u)$. Similarly, $Perm^{'C_{v}}(v) > Perm^{C_{v}}(v)$.

\textbf{Case D.3.2: $C_{u}$ is responsible for $E_{max}(v)$ but $C_{v}$ is not responsible for $E_{max}(u)$:}
The new $D(v)$ i.e.,$D^{'}(v) < D(v)$ and the new $E_{max}(v)$ i.e., $E_{max}^{'}(v) < E_{max}(v)$ because one edge goes away. $I(v)$ remains the same. Therefore, the new permanence of $v$ i.e., $Perm^{'C_{v}}(v) > Perm^{C_{v}}(v)$. Also, $Perm^{'C_{u}}(u) > Perm^{C_{u}}(u)$.

\textbf{Case D.3.3: $C_{u}$ and $C_{v}$ do not influence $E_{max}(v)$ and $E_{max}(u)$, respectively:}
The new $d(u)$ i.e., $d^{'}(u) < d(u)$ and $E_{max}(u)$, $I(u)$ remain the same. Therefore, the new permanence of $u$ i.e., $Perm^{'C_{u}}(u) > Perm^{C_u}(u)$. Similarly, $Perm^{'C_{v}}(v) > Perm^{C_{v}}(v)$.

\textbf{Case D.3.4: Both $C_{u}$ and $C_{v}$ influence  $E_{max}(v)$ and $E_{max}(u)$, respectively:}
The new $E_{max}(u)$ i.e.,$E_{max}^{'}(u) < E_{max}(u)$ and the new degree decreases by 1 i.e., $d^{'}(u) < d(u)$. Therefore, the new permanence $Perm^{'C_{u}}(u) > Perm^{C_{u}}(u)$.
Similarly, the new $E_{max}(v)$ i.e., $E_{max}^{'}(v) < E_{max}(v)$ and $d^{'}(v) < d(v)$. Therefore, $Perm^{'C_{v}}(v) > Perm^{C_{v}}(v)$.

\textbf{Case D.4: Both $u$ and $v$ belong to the same community i.e., intra-community link:} Assume that both $u$ and $v$ belong to community $C$. After the edge between $u$ and $v$ is deleted,  permanence of both the nodes decreases as shown in Proposition \ref{prop4}. Therefore, $C$ may split. Algorithm 7 handles this case.

\begin{proposition}\label{prop4}
Deleting an intra-community edge between nodes $u$ and $v$ decreases the permanence value of the two nodes.
\end{proposition}
\if{0}
\begin{proof}
Let $P_a(u) = \frac{I(u)}{E_{max}(u)} \frac{1}{d(u)}$ and $P_b(u) = 1 - C_{in}(u)$.\\
Let's suppose that an edge connecting the nodes $u$ and $v$, which are both members of a community $C$, is removed. \\
For the node $u$, let $Perm^{C}(u)$ be the permanence before deleting the edge and let $Perm^{'C}(u)$ be its permanence afterwards.
For the proposition to be true, we have to prove,\\
$Perm^{C}(u) - Perm^{'C}(u) < 0$ i.e., $\Delta Perm^{C}(u)< 0$.\\
$\Delta Perm^{C}(u) = \frac {I(u)}{d(u)} - \frac{I(u)-1}{d(u)-1} + \frac {x+I(u)-1}{I(u)(I(u)-1)} - \frac{I(u)-1}{(I(u)-1)(I(u)-2)}$. \\
Simplified further, $\frac{xI(u)-2x+2-2I(u)}{I(u)(I(u)-1)(I(u)-2)}-\frac{d(u) - I(u)}{d(u)(d(u)-1)} > 0$. Here, $x$ represents the number of links between the node $u$ and its other internal neighbors, apart from $v$.\\
The denominator of the above expression is greater than $0$ as all terms are positive, so we solve for the numerator we get, \\
$d(u)(d(u)-1)(xI(u)-2x+2-2I(u))-I(u)(d(u)-I(u))(I(u)-1)(I(u)-2)$.\\ On solving further, we get\\
$I(u)^{4}+I(u)^{3}(3-d(u))+I(u)^{2}(3d(u)-2)+I(u)(xd(u)(d(u)-1)-2d(u)^{2})+2d(u)(d(u)-xd(u)+x-1) > 0$. \\
Now, $I(u)^{4}$ is always positive. \\
$I(u)^{3}(3-d(u)) > 0$ if $d(u)<3$.\\
$I(u)^{2}(3d(u)-2) > 0$ if $d(u) \neq 0$.\\
$I(u)(xd(u)(d(u)-1)-2d(u)^{2}) > 0$,  if $d(u) = 0$. \\
$2d(u)(d(u)-xd(u)+x-1) > 0$, if $d(u) = 0,1$. \\
From the above derivation we see that there is no value of d(u) for which all the expressions are positive. Hence, $Perm^{C}(u) - Perm^{'C}(u) > 0$.
\end{proof}
\fi
See Supplementary \cite{si} for the proof.


\if{0}
\begin{algorithm}
  \textbf{Input:} Graph G (V,E), Edge e($u, v$) to be deleted; Current structure $C_t$.
\textbf{Output:} Updated community structure $C_{t+1}$ \\
  $C_u \leftarrow Comm(G,u)$;\\
  $C_v \leftarrow Comm(G,v)$;\\
  \SetAlgoLined
  \textbf{if} $C_u == C_v$ \textbf{then}\\
  \Indp $C_{t+1} \leftarrow$ IntraEdgeDeletion($G$, $e$, $C_t$)  //Algorithm 7\\
  \Indm \textbf{else}\\
  \Indp $C_{t+1} \leftarrow$ InterEdgeDeletion($G$, $e$, $C_t$)  //Algorithm 8\\
 
\Indm \textbf{return} $C_{t+1}$; \\
  \caption{EdgeDeletion}
\end{algorithm}

\begin{algorithm}
  \textbf{Input:} Graph G (V,E), Edge {$u, v$} to be deleted; Current structure $C_t$.

  \textbf{Output:} Updated community structure $C_{t+1}$ \\
  \SetAlgoLined
  $visited$ $\leftarrow$ $null$, $queue$ $\leftarrow$ $null$; \\
  $C_{uv} \leftarrow Comm(G,u)$ \\
  $G(V,E)$ $\leftarrow$ $G(V,E)$ - ($u,v$) \\
  Append $u$ to queue; \\
  Add head($queue$) to $visited$;\\
  \textbf{while} (length($queue$) $>$ 0) \textbf{do} \\
 
        \Indp \textbf{for each} unvisited node $i$ in $queue$ \textbf{do} \\
            \Indp $C_u$ $\leftarrow$ $C_u$ $\cup$ head($queue$);\\
      $visited$ $\leftarrow$ $visited$ $\cup$ head($queue$); \\
      \Indm \Indm \textbf{for each} node $q$ in $queue$ \textbf{do}\\
      \Indp Add unvisited internal neighbors of $q$ to $queue$; \\
      \Indp Remove $q$;\\
      \Indm \Indm Follow step 5 to 13 to obtain $C_v;$\\
      \textbf{if} $Perm(C_{uv})<Perm(C_u)+Perm(C_v)$ \textbf{do}\\
      \Indp $C_{t+1}=C_t - C_{uv} \cup C_u \cup C_v$;\\
      \Indm else\\
      \Indp $C_{t+1}=C_t$;\\
  \Indm \textbf{return} $C_{t+1}$; \\
  \caption{IntraEdgeDeletion}
\end{algorithm}
\begin{algorithm}
  \textbf{Input:} Graph G (V,E), Edge e($u, v$) to be deleted; Current structure $C_t$.
\textbf{Output:} Updated community structure $C_{t+1}$ \\
  $C_u \leftarrow Comm(G,u)$;\\
  $C_v \leftarrow Comm(G,v)$;\\
  \SetAlgoLined
  $G(V,E)$ $\leftarrow$ $G(V,E)$ - ($u,v$) \\
 $C_{t+1} \leftarrow C_{t}$\\
 \textbf{return} $C_{t+1}$; \\
  \caption{InterEdgeDeletion}
\end{algorithm}
\fi

\begin{figure}[!t]
\centering
\includegraphics[width=\columnwidth]{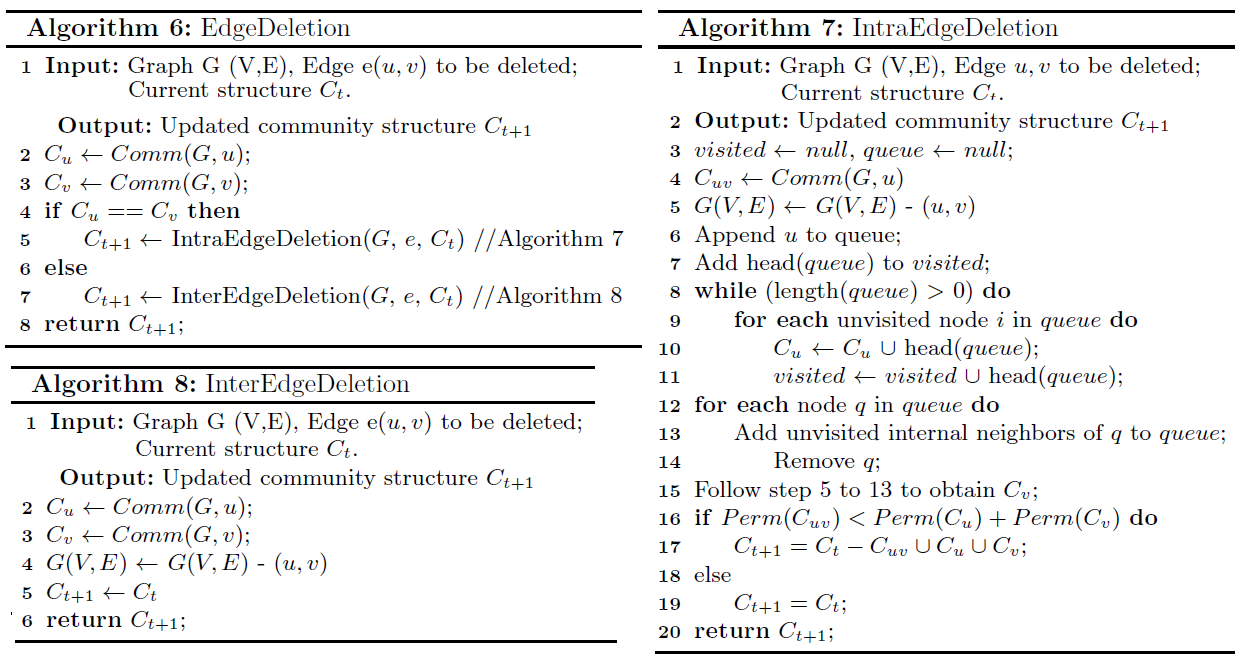}
\vspace{-10mm}
\end{figure}

\noindent\fbox{%
    \parbox{\textwidth}{%
The time complexity of \name~is $\mathcal{O}(E)$ (see Supplementary \cite{si} for the detailed complexity analysis).}}

\if{0}
\vspace{-3mm}
\subsubsection{\underline{Time complexity}:}

\name~ consists of 4 major subroutines -- $NodeAddition$, $NodeDeletion$, $EdgeAddition$ and $EdgeDeletion$. $Node Addition$ and $NodeDeletion$ are represented in the form of $EdgeAddition$ and $EdgeDeletion$ respectively. Therefore, the time complexity of \name~ will be the maximum of $EdgeAddition$ and $EdgeDeletion$.
The time complexity of $EdgeAddition$ and $EdgeDeletion$ is $O(V+E)$. 
Therefore, the total time complexity of \name~ is $O(V+E)$ (see Supplementary \cite{si} for the detailed discussion).
\fi

\vspace{-5mm}
\section{Experimental results}
\vspace{-3mm}
In this section, we start by briefly describing the datasets and baseline methods, followed by the detailed experimental results.

\vspace{-5mm}
\subsection{Datasets} 
\vspace{-3mm}
We perform our experiments on two types of networks: synthetic networks and real-world networks. \\ 
\textbf{\underline{Synthetic networks}:} To generate synthetic networks and their ground-truth communities, we use the dynamic LFR benchmark model\footnote{\url{http://mlg.ucd.ie/snam/}} \cite{Greene:2010}. It allows  users to specify different parameters -- number of nodes ($N$), mixing coefficient ($\mu$) which controls the ratio of external neighbors of a node to its degree, the average ($k$) and maximum degree ($k_{max}$), and the number of time-stamps $s$ to generate the dynamic network. Here, we vary $N$ from $500$ to $3500$, $s$ from $10$ to $30$, $\mu$ from $0.10$ to $0.80$. The default values are considered for the rest of the parameters. However, unless otherwise mentioned, the default LFR network is generated with the following parameter setting: $N=1000$, $s=20$, $\mu=0.2$.\\
\textbf{\underline{Real-world networks}:} Four real-world dynamic networks are used whose ground-truth communities are known to us: (i) \textbf{Cumulative co-authorship network} (Coauth-C) \cite{ChakrabortySTGM13,0002SGM14}, (ii) \textbf{Non-cumulative co-authorship network} (Coauth-N) \cite{Chakraborty:2014},
(iii) \textbf{2011 High school dynamic contact networks} (HS-11)\footref{note1}, (iv) \textbf{2012 High school dynamic contact networks} (HS-12)\footnote{\label{note1}\url{http://www.sociopatterns.org/}}, 
(iii) \textbf{Primary school contact networks} (PS)\footref{note1}, 
and (iv) \textbf{Contact network in a workplace} (CW)\footref{note1}.
Table \ref{tab:dataset} presents statistics of the datasets (see more details in supplementary \cite{si}).

\vspace{-5mm}
\begin{table}
\caption{Description of the real-world networks (notation: $N$ ($E$):  \# of unique nodes (edges), $\bar{N}$ ($\bar{E}$): avg. \# of nodes (edges) per time-stamp, $\bar{C}$: avg. \# of communities per time-stamp, $s$: \# of time-stamps).}\label{tab:dataset}
\vspace{-5mm}
\begin{center}
\scalebox{0.9}{
\begin{tabular}{ l|c|c|c|c|c|c|c }
\hline
  {\bf Network}  & {\bf N ($\bar{N}$)} & {\bf Node-type} &  {\bf E ($\bar{E}$)} & {\bf Edge-type} & {\bf $\bar{C}$} & {\bf Community-type} & {\bf s} \\\hline\hline
Coauth-C 	&	 708497(41676)	& Author &	1166376(68610)  & Coauthorship	&24	 & Research area	&	17 \\
Coauth-N 	&	708497(41676) 	& Author &	1166376(68610)& Coauthorship	&24	 & Research area	&	17 \\
HS'11 	&	126(18)	& Student &	1710(244)	& Contact &	3  &	Class & 7 \\
HS'12 	&	180(22.5)	& Student &	2220(225)	& Contact &	5  &	Class &  8\\

PS 	&	242(47)	& Student &	 77602(323) & Contact	&	5 & Class	&	6 \\
CW 	&	145(18)	& Individual &	1193(149)	& Contact &	5	& Department &	8\\
\hline
\end{tabular}}

\end{center}
\vspace{-12mm}
\end{table}

\vspace{-5mm}
\subsection{Baseline methods}
\vspace{-2mm}
We use the following state-of-the-art  dynamic community detection methods to compare with \name: 
(i) {\bf Quick Community Adaptation (QCA):} This framework uses a modularity-based approach for dynamic community detection \cite{5935045}; (ii) \textbf{Fast Community Detection for Dynamic Complex Networks (FCDDCN):} This is a community detection method for real-time dynamic networks. Modularity is optimized using heuristic search \cite{Bansal2011}; (iii) \textbf{GreMod:} It is an incremental algorithm that performs per-determined actions for every edge change to maximize modularity \cite{ShangLXCMFW14}; (iv) \textbf{Learning-based Targeted Revision (LBTR):} It  uses machine learning classifiers to predict the vertices that need to be inspected for community assignment revision \cite{shang2016targeted}. We used the source code of these algorithms, released by the authors.

\vspace{-5mm}
\subsection{Comparative evaluation}
\vspace{-3mm}
We compare  the  obtained community structure with  a  given  ground-truth  structure based on the following metrics: Normalized Mutual Information (NMI) and Adjusted Rand Index (ARI). The value of NMI ({\em resp.} ARI) ranges from $0$ ({\em resp.} -1) (no match) to 1 (perfect match). We design two experimental setups to perform a thorough comparative analysis.

\begin{figure}[!t]
\includegraphics[width=\linewidth, height= 4cm]{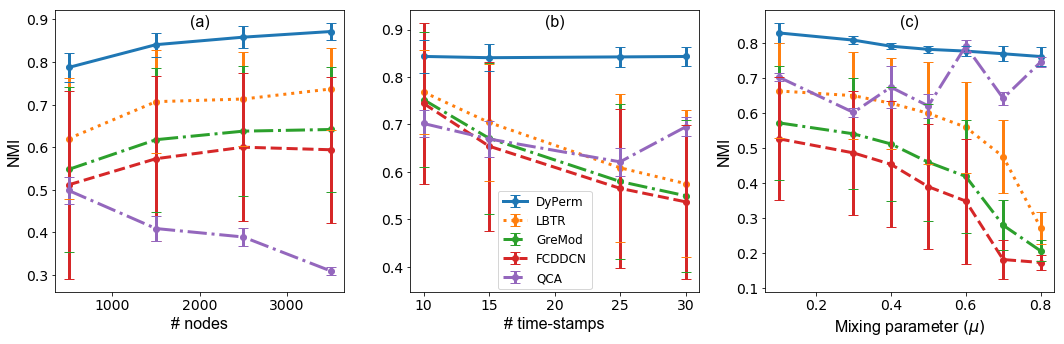}
\vspace{-10mm}
  \caption{Accuracy (average NMI and its standard deviation across different time-stamps for each network) of the competing methods with the change of LFR parameters for experimental setup I (similar pattern is observed for ARI, see Supplementary \cite{si}).  }\label{fig:image1}
  \vspace{-5mm}
 \end{figure}

\noindent \textbf{Experimental setup I: Running best static community detection method to obtain base communities. }
In order to obtain the base community structure $C_0$ for \name, we run  MaxPerm (a permanence maximization algorithm for static networks) \cite{Chakraborty:2014} on the initial snapshot of the network. Since all the baseline methods maximize modularity, we run Louvain algorithm (a modularity maximization algorithm for static network) \cite{louvain} on the initial snapshot. In each time-stamp, we compare the output of each competing method with the ground-truth community structure and report the average accuracy and the standard deviation.

\vspace{-5mm}
\begin{table}[!th]
\centering
\caption{Accuracy (avg. NMI and ARI) of the competing methods on the default LFR and real-world networks for experimental setup I. Top results are in bold-face.}
\label{accuracy1}
\scalebox{0.85}{
\begin{tabular}{l|cc|cc|cc|cc|cc}
\hline
\multirow{2}{*}{{\bf Dataset}} & \multicolumn{2}{c|}{{\bf QCA} }& \multicolumn{2}{c|}{{\bf LBTR}} & \multicolumn{2}{c|}{{\bf GreMod}} & \multicolumn{2}{c|}{{\bf FCDDCN}} & \multicolumn{2}{c}{{\bf \name}} \\\cline{2-11}
 & {\bf NMI}  & {\bf ARI}  & {\bf NMI}  & {\bf ARI} & {\bf NMI} & {\bf ARI} & {\bf NMI}  & {\bf ARI}  & {\bf NMI}  & {\bf ARI} \\ \hline\hline
 LFR (default)   & 0.55         &      0.41       &     0.65       &       0.41    &        0.57       &      0.34    &      0.53        &      0.36         &       {\bf 0.81}           &     {\bf 0.54}  \\ 
 Coauth-C     &    0.37   &   0.03  &    0.04    &   0.01 &    0.05 & 0.08  & 0.05 & 0.03 & {\bf 0.49} &  {\bf 0.11} \\  
 Coauth-N      &   0.39   &  0.04 &    0.04    & 0.01 &   0.04   &   0.05 & 0.03  & 0.02  & {\bf 0.48}  &  {\bf 0.11} \\  
 HS'11      &    0.39  & 0.02  & 0.04   &  0.06  &  0.04  & 0.06  & 0.04  &  0.05   &  {\bf 0.59}  &  {\bf 0.13}    \\  
 {HS'12}      &    0.43 & 0.19  & 0.02   &  0.05  &  0.02 & 0.05  & 0.02   &  0.04  & {\bf 0.56}  &  {\bf 0.24}    \\  
 {PS}      &    0.39  & 0.14  & 0.04   &  0.02  &  0.04  & 0.02 & 0.04   &  0.01   & {\bf 0.53} &  {\bf 0.25}   \\  
 {CW}      &    0.41 & 0.01  & 0.02    &  0.07  &  0.03  & 0.01  & 0.03    &  0.03   & {\bf 0.52}  &  {\bf 0.09}    \\ \hline
\end{tabular}}
\vspace{-10mm}
\end{table}

Figure \ref{fig:image1} shows the NMI value (and its standard deviation) of the competing methods with the change in different parameters  of the LFR networks (see Supplementary \cite{si} for the same plot w.r.t ARI). We observe that the NMI value of \name~is consistently higher than those of the baseline methods irrespective of any LFR parameters. \name~outperforms the best baseline method (LBTR) by 20.6\%, 26.74\%, 35.75\% on average with the increase of the number of nodes, time-stamps and $\mu$ respectively, which is significant
according to the $t$-test with 95\% confidence interval. The standard deviation of \name~is also less compared to that of LBTR, indicating that \name~is consistent in producing accurate community structure across different time-stamps of a network.
Table \ref{accuracy1} shows the results of the competing methods on the default LFR and real-world networks. Once again, we observe a significant gain in the performance of \name~compared to the other baselines, specially on the real-world networks. QCA turns out to be the bast baseline method for real-world networks. \name~outperforms QCA by 35.20\% and 275.4\% in terms of NMI and ARI  respectively, averaged across all the networks. \\

\noindent\textbf{Experimental setup II: Using ground-truth community structure as base communities.} We would like to reiterate that all the dynamic community detection methods are highly dependent on the base community structure. The noise in the detection of the base community structure may propagate to the next stage of the algorithm and affect the overall performance. Therefore, one may argue that the  baseline methods seem to be incompetent (as observed in Table \ref{accuracy1})  due to the inefficiency of the static community detection method applied on the initial snapshot, not due to the problem in dynamic community detection method itself. To verify this argument further, we use the ground-truth community structure of the initial network as the base community structure. This ensures that the base community structure is completely accurate. Following this, we run each competing dynamic method on the remaining snapshots and measure the accuracy.

\begin{figure}[!t]
\includegraphics[width=\linewidth]{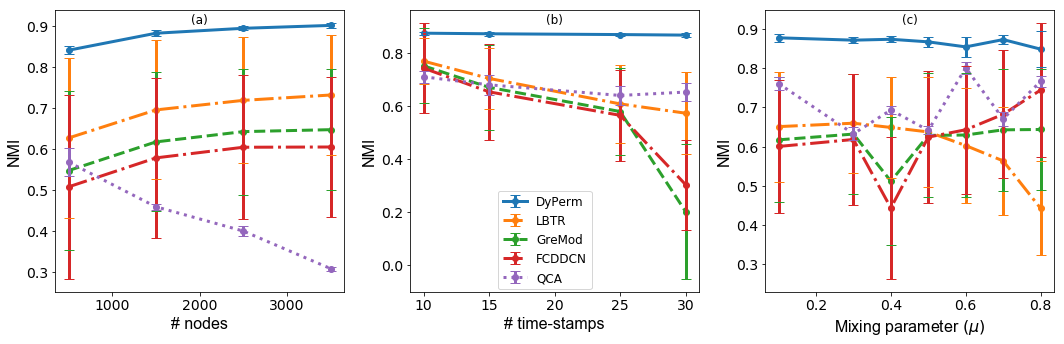}
\vspace{-10mm}
  \caption{Accuracy (average NMI and its standard deviation across different time-stamps for each network) of the competing methods with the change of LFR parameters for experimental setup II (similar pattern is observed for ARI, see Supplementary \cite{si}).}\label{fig:image2}
  \vspace{-5mm}
 \end{figure}

Figure \ref{fig:image2} shows the NMI value (and its standard deviation) of the competing methods with the change in different LFR parameters (see Supplementary \cite{si} for the same plot w.r.t. ARI). Once again, we observe similar pattern -- \name~significantly outperforms all other baseline methods. However, here both LBTR and QCA seem to be quite competitive.  \name~beats LBTR by 27.11\%, 30.88\% and 45.03\%, and QCA by 111.5\%, 29.17\% and 23.89\% with the increase of the number of nodes, time-stamps and $\mu$ respectively, averaged over all the time-stamps. These improvements are significant
according to the $t$-test with 95\% confidence interval. Table \ref{accuracy2} shows the accuracy of the competing methods on the LFR and different real-world networks for experimental setup II. We again observe a significant improvement of the performance of \name~comapred to the baselines. This implies that irrespective of the community detection method used on the initial snapshot of the network, our method always outperforms other baselines.

\begin{table}[!h]
\centering
\caption{Accuracy (avg. NMI and ARI) of the competing methods on the default LFR and real-world networks for experimental setup II. Top results are in bold-face.}
\label{accuracy2}
\scalebox{0.85}{
\begin{tabular}{l|cc|cc|cc|cc|cc}
\hline
\multirow{2}{*}{{\bf Dataset}} & \multicolumn{2}{c|}{{\bf QCA} }& \multicolumn{2}{c|}{{\bf LBTR}} & \multicolumn{2}{c|}{{\bf GreMod}} & \multicolumn{2}{c|}{{\bf FCDDCN}} & \multicolumn{2}{c}{{\bf \name}} \\\cline{2-11}
 & {\bf NMI}  & {\bf ARI}  & {\bf NMI}  & {\bf ARI} & {\bf NMI} & {\bf ARI} & {\bf NMI}  & {\bf ARI}  & {\bf NMI}  & {\bf ARI} \\ \hline\hline

LFR (default) & 0.57  	& 0.47   & 0.76   & 0.55  &	0.62   & 0.28   & 0.55   & 0.41   & {\bf 0.87}   & {\bf 0.58}  \\  			
Coauth-C     &    0.13     &   0.03   &    0.03      &   0.02  &    0.04    & 0.08    & 0.05   & 0.05   & {\bf 0.53}   &  {\bf 0.12}  \\  										  											
Coau-N      &   0.10     &  0.05  &    0.02      & 0.04   &   0.04   &   0.01 & 0.05   & 0.01  & {\bf 0.53}   &  {\bf 0.12}  \\  										
						
HS'11      &    0.08   & 0.05   & 0.04     &  0.06   &  0.04   & 0.06   & 0.03    &  0.05    & {\bf 0.52}   &  {\bf 0.10}     \\  						
HS'12     &    0.02   &0.09   & 0.02    &  0.05   &  0.03   & 0.02   & 0.03     &  0. 03   & {\bf 0.60}   &  {\bf 0.22}     \\  																								
PS    &    0.03  & 0.08   & 0.04     &  0.02   &  0.04   & 0.02   & 0.04    &  0.02    & {\bf 0.53}   &  {\bf 0.23}     \\  

CW      &    0.03   &0.02   & 0.02    &  0.03  &  0.04   & 0.06    & 0.05    &  0.04    & {\bf 0.57}   &  {\bf 0.09}     \\  \hline

\end{tabular}}
\vspace{-3mm}
\end{table}

Interesting, while comparing Tables \ref{accuracy1} and \ref{accuracy2}, we notice that the performance of the baselines does not improve much considering ground-truth as base community structure, specially for the real-world networks. However, \name~seems to achieve a significance performance gain in most cases -- 4.41\% and 1.23\% in terms of NMI and  ARI, averaged over all the datasets. This implies that with a better initialization of the community structure, \name~can achieve even better performance.

\vspace{-5mm}
\subsection{Run-time analysis}
\vspace{-3mm}
The motivation behind designing any dynamic community detection method is that it runs faster than a static method. In case of static community detection method, after every change in the network structure, we need to consider the resultant network as a whole and run the method; whereas an incremental dynamic community detection method only considers those parts of the network, which have been changed and modified the earlier community structure accordingly without running the entire method from the scratch. In Table \ref{runtime}(a), we report the runtime of MaxPerm and \name, the static and dynamic community detection methods which maximize permanence, respectively. \name~seems to be 10 times faster than MaxPerm, averaged over all the real-world  datasets. Maximum gain ($15$ times faster) is observed on Coauthor-N network. This result provides enough motivation to design an efficient dynamic community detection method. Note that we can not compare the runtime of other competing methods as the source codes were written in different languages. We therefore compare the theoretical time complexity of these methods in Table \ref{runtime}(b).

\begin{table}[!t]
\begin{center}
\caption{(a) Runtime (in minutes) of MaxPerm and \name~for different real-world networks. (b) Time complexity of the competing  methods ($N$: \# of nodes, $E$: \# of edges, $d$: avg. degree of nodes).}
\label{runtime}
\vspace{-3mm}
\scalebox{0.9}{
\begin{tabular}{ l|c|c|c|c|c|c }
\multicolumn{7}{c}{(a)}\\
\hline
\multirow{2}{*}{{\bf Method}}  & \multicolumn{6}{c}{{\bf Runtime (in minutes) for different datasets}}\\\cline{2-7}
  & Coauth-C & Coauth-N & HS'11 & HS'12 & PS & CW\\\hline\hline
MaxPerm & 3,420 & 3,020 & 180 & 192 & 17 & 40 \\
\name~ & 300 & 204 & 45 & 48 & 1.5 & 5 \\\hline
\end{tabular}}
\scalebox{0.9}{
\begin{tabular}{c|c|c|c|c}

\multicolumn{5}{c}{(b)}\\\hline
QCA & LBTR & GreMod & FCDDCN & \name \\\hline\hline
$\mathcal{O}$($E^2$) & $\mathcal{O}$($E$) & $\mathcal{O}$($E$) & $\mathcal{O}$($Ed\log N$) & $\mathcal{O}$($E$) \\
\hline

\end{tabular}}

\end{center}
\vspace{-10mm}
\end{table}

\vspace{-5mm}
\section{Conclusion}
\vspace{-5mm}
In this paper, we proposed \name, a novel dynamic community detection method that maximizes permanence (a local community scoring metric) in every snapshot of the network to detect the community structure. \name~significantly outperformed four state-of-the-art baselines on both synthetic and real-world networks -- we observed a gain in NMI up to 35\% compared to the best baseline method. Moreover, \name~truned out to be extremely faster than its static counterpart (MaxPerm), achieving up to 15 times speedup. We also presented theoretical analysis to show how/why minimum local changes in the community structure leads to permanence maximization in dynamic networks.  We have also released the anonymized code and datasets for the sake of reproducibility at \url{https://tinyurl.com/dyperm-code}.

%
%

\bibliographystyle{splncs03.bst}

\end{document}